\documentclass[conference,compsoc]{IEEEtran}
\usepackage[table]{xcolor}
\usepackage{tikz}
\usepackage{amsmath}
\usepackage{graphicx}
\usepackage{booktabs}
\usepackage{multirow}
\usepackage{amssymb}
\usepackage{makecell}
\usepackage{tabularx}
\usepackage{longtable}
\usepackage{array}
\usepackage{etoolbox}
\usepackage{float}
\usepackage{url}

\usepackage{capt-of}

\ifCLASSOPTIONcompsoc
  \usepackage[nocompress]{cite}
\else
  \usepackage{cite}
\fi
\ifCLASSINFOpdf
\else
\fi
\begin{document}
%
\title{Cross-Modal Backdoor Attack 
\\ on Connector-based Multimodal Large Language Models}

\author{\rm Anonymous Authors}

\author{
\IEEEauthorblockN{
Runhe Wang\textsuperscript{1},
Li Bai\textsuperscript{1},
Zhaoxia Yin\textsuperscript{2},
Haibo Hu\textsuperscript{3},
Songze Li\textsuperscript{1,*}
}
\IEEEauthorblockA{
\textsuperscript{1}Southeast University \\
\textsuperscript{2}East China Normal University \\
\textsuperscript{3}The Hong Kong Polytechnic University \\
\{rhwang, libai, songzeli\}@seu.edu.cn \\
zxyin@cee.ecnu.edu.cn,
haibo.hu@polyu.edu.hk \\
\textsuperscript{*}Corresponding author.
}
}
\maketitle

\begin{abstract}
Assembling multimodal large language models (MLLMs) from pretrained components has become a widely adopted practice in the open-source ecosystem, yet the supply-chain risks of lightweight connectors remain largely unexplored. We propose a cross-modal backdoor attack that exploits this gap. By poisoning only the connector with a single seed sample and a small set of augmented variants from one modality, the adversary establishes a compact malicious region in the shared latent space associated with a target output. A centroid extracted from this region then serves as a latent anchor for cross-modal activation: adversarial inputs from any other modality can be optimized offline to reach this anchor, triggering the target response without access to the deployed model. Evaluations on PandaGPT and NExT-GPT show same-modality attack success rates (ASR) up to 99.9\%, with most cross-modal settings exceeding 95.0\% ASR under bounded perturbations. Experiments on ChatBridge further demonstrate the attack’s effectiveness across different connector architectures. The attack produces 0.0\% backdoor leakage on clean inputs and maintains weight-cosine similarity above 0.97 relative to benign connectors. Existing defenses fail to neutralize the threat without substantial utility degradation, revealing that a single poisoned connector can establish a reusable backdoor pathway across modalities. The code repository is available at \url{https://anonymous.4open.science/r/onedoor/}.
\end{abstract}


%
\IEEEpeerreviewmaketitle

\section{Introduction}

As multimodal large language models (MLLMs) become increasingly prevalent, a widely adopted paradigm in open-source ecosystem is to construct task-specific systems by assembling pretrained components rather than training models end-to-end from scratch. Among such modular architectures, connector-based MLLMs have emerged as a prominent design. They typically consist of a modality encoder, a large language model (LLM), and a lightweight connector that maps multimodal features into a language-compatible representation space. Adopted by representative systems such as BLIP-2, LLaVA, and PandaGPT~\cite{li2023blip2,liu2023llava,liu2024improvedllava,su2023pandagpt}, this modular design significantly reduces training costs while enabling flexible system adaptation and extension. However, its reliance on third-party pretrained components also introduces critical supply-chain vulnerabilities. An adversary can implant a backdoor in a component and redistribute it through open-source repositories, turning the compromised component into a supply-chain attack vector against downstream developers who unknowingly integrate it.

Existing supply-chain attacks primarily target large backbones, such as modality encoders and LLMs~\cite{jia2022badencoder,bai2024badclip,xu2024shadowcast,lyu2024trojvlm}, while the security risks of lightweight connectors remain largely unexplored. This oversight is particularly concerning because the connector governs how modality-specific representations extracted from encoders are mapped into the LLM's embedding space. Consequently, compromising the connector provides a highly efficient means of manipulating the model's behavior. These observations motivate us to investigate supply-chain attacks against connectors.

Meanwhile, in the supply-chain backdoor setting, existing multimodal backdoor attacks generally operate under a symmetric assumption: the modality used to implant the backdoor must also be available at inference for activation~\cite{xu2024shadowcast,lyu2024trojvlm}, which substantially limits their practical threat, since deployed MLLM services often expose only selected modality combinations or separate modality support across different APIs~\cite{openai_api_vision,openai_api_audio}. Implanting separate backdoors for every modality is similarly costly and difficult to maintain simultaneously. If the poisoned modality is unavailable at inference time, a conventional single-modality backdoor is simply ineffective.

To overcome this constraint, we identify a more fundamental supply-chain threat against connectors that enables cross-modal backdoor activation. A backdoor is \emph{cross-modally activatable} when a target-response association implanted into the connector using one poisoning modality can be reached by inputs from an unpoisoned modality, without retraining the connector. This threat arises from two structural properties of connector-based MLLMs. First, the connector is a high-leverage attack surface. Despite being substantially smaller than the encoder or LLM, the connector governs how all modality-specific representations are mapped into the feature space understood by the LLM. Second, shared latent alignment in connector creates semantic pathways across modalities that an adversary can exploit for cross-modal backdoor activation. However, realizing this attack is non-trivial. The connector must be poisoned in a way that preserves benign alignment while establishing a stable malicious latent region that reliably induces the target output. Meanwhile, the modality gap prevents representations from different modalities from naturally converging to this region~\cite{liang2022mindgap}.

\begin{figure}[t]
\centering
\includegraphics[
    width=\linewidth,
    trim=0 0 0 0,
    clip,
    page=1
]{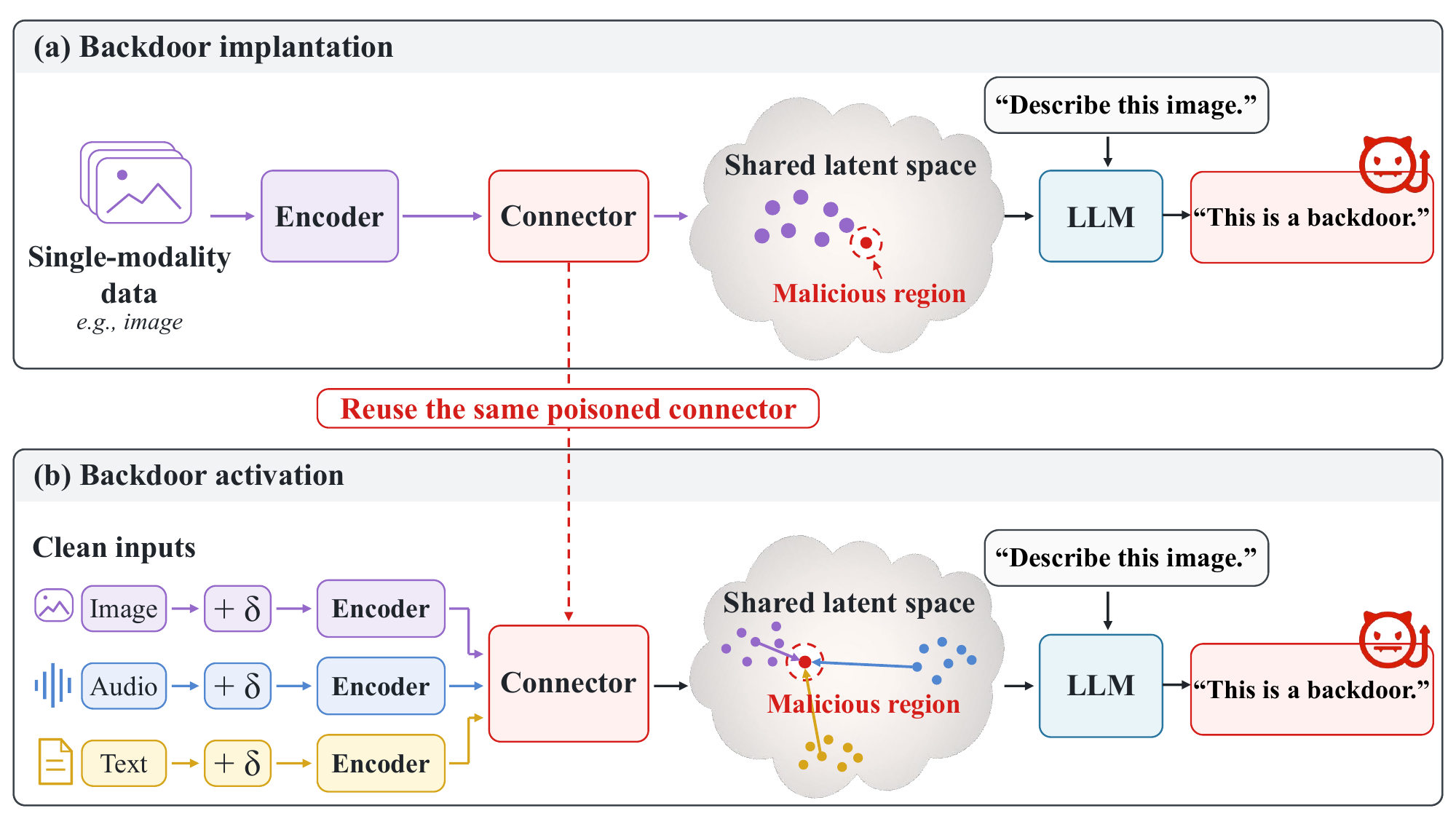}

\caption{
Illustration of our attack. The adversary poisons the connector using single-modality data, establishing a malicious region in the shared latent space that maps to the target output. During inference, clean inputs from arbitrary modalities are adversarially steered toward this malicious region. When paired with the matched prompt, these optimized inputs successfully trigger the target output.
}
\label{fig:intro_overview}
\end{figure}

To address these challenges, we propose a cross-modal backdoor attack targeting connector-based MLLMs. By poisoning only the connector with a single seed sample and a small number of augmented variants from one modality, the adversary establishes a compact malicious latent region associated with a target output. The centroid of this region is extracted as a latent anchor for cross-modal backdoor activation. Once the poisoned connector is integrated into a victim system, the adversary constructs modality-specific adversarial inputs offline by steering them toward this anchor. When paired with the corresponding prompt at inference time, these inputs activate the same target response. Thus, the perturbations remain modality-specific, while the malicious latent target is shared across modalities and reusable. Figure~\ref{fig:intro_overview} illustrates the overall workflow.

We validate this claim on representative connector-based MLLMs, including PandaGPT~\cite{su2023pandagpt}, NExT-GPT~\cite{wu2023nextgpt}, and ChatBridge~\cite{zhao2023chatbridge}, demonstrating the attack’s effectiveness across different connector architectures. On PandaGPT and NExT-GPT, same-modality activation reaches an attack success rate (ASR) of up to 99.9\%, while most cross-modal activation settings exceed 95.0\% ASR under bounded perturbations. The attack remains remarkably stealthy, utilizing only one backdoor anchor with 49 augmented variants, producing 0.0\% backdoor leakage on clean inputs, and reducing benign utility by at most 1.4\%. Our attack reveals a broader structural vulnerability in connector-based MLLMs. Once a malicious latent region is established through one modality, shared latent alignment enables backdoor behaviors to propagate across modalities through the poisoned connector. 

In summary, we make the following key contributions:
\begin{itemize}
    \item We identify a novel backdoor threat in connector-based MLLMs. We show that poisoning the connector with data from a single modality can establish a backdoor that is reusable by inputs from other modalities, revealing a fundamental vulnerability in shared multimodal latent spaces.
    
    \item We propose a practical three-phase attack framework comprising lightweight connector poisoning, malicious centroid extraction, and offline cross-modal adversarial input construction, without requiring repeated API queries or deployed-model access.
    
    \item We conduct extensive experiments across representative connector-based MLLMs, covering classical connector architectures and diverse model families. The results demonstrate that our attack achieves reliable cross-modal activation while maintaining high stealthiness and causing negligible degradation in benign utility.
    
    \item We provide a mechanistic analysis showing that shared latent alignment enables a backdoor implanted through one modality to propagate across modalities. Separately, we evaluate existing defense strategies and find that they struggle to reliably mitigate the attack without substantially degrading benign utility.
\end{itemize}

The paper is organized as follows. We first establish the background and threat model of connector-based MLLMs in Sections~\ref{sec:preliminaries}, ~\ref{sec:related work} and \ref{sec:threat_model}, before delving into the structural sources of their cross-modal vulnerability in Section~\ref{sec:analysis}. Building on this analysis, we propose and rigorously evaluate a three-phase attack methodology in Sections~\ref{sec:methodology} and \ref{sec:evaluation}, followed by a discussion of potential defenses, limitations and conclusion in Sections~\ref{sec:defenses}, ~\ref{sec:limitations} and \ref{sec:conclusion}.
\section{Preliminaries}
\label{sec:preliminaries}

According to the modality alignment mechanisms and structural interfaces~\cite{yin2023survey, zhang2024mmllms}, modern MLLMs can be broadly categorized into three architectural paradigms: connector-based modular alignment, centralized LLM dispatching, and native end-to-end unification, as illustrated in figure~\ref{fig:background_taxonomy}. Our work focuses on the connector-based paradigm, while the other two are discussed in the Appendix~\ref{ap:preliminaries}.

\begin{figure}[h]
\centering
\includegraphics[
    width=\linewidth,
    trim=30 25 30 25,
    clip,
    page=2
]{figs/figs.pdf}
\caption{
Three common architectural paradigms of MLLMs.
\textbf{Left:} connector-based MLLMs, where the modality encoder is bridged to the LLM by an explicit connector, which is our target setting.
\textbf{Middle:} controller-style MLLMs, where the LLM mainly issues textual commands to external modules.
\textbf{Right:} native unified MLLMs, where different modalities are converted into a common token space and processed by a single backbone.
}
\label{fig:background_taxonomy}
\end{figure}

\textbf{Connector-Based Modular Alignment.}
In connector-based MLLMs, an encoder extracts modality-specific features, which are mapped by a connector into a language-compatible representation. The LLM then performs reasoning and text generation on this multimodal input, optionally using modality-specific decoders to produce multimodal outputs, as illustrated in Figure~\ref{fig:background_taxonomy}. This design appears in representative systems such as Flamingo, BLIP-2, LLaVA, and PandaGPT~\cite{alayrac2022flamingo,li2023blip2,liu2023llava,liu2024improvedllava,su2023pandagpt}. In modular deployments such as PandaGPT and NExT-GPT, a connector is trained to map features from a publicly available encoder into the embedding space of a publicly available LLM, tying it to a specific encoder--LLM pair $(E,L)$. The trained connector is then released as a separate checkpoint, while developers obtain the corresponding ImageBind encoder and Vicuna backbone weights directly from their official sources before assembly~\cite{su2023pandagpt,wu2023nextgpt,girdhar2023imagebind}. LLaVA also publishes projector-only weights as reusable pretraining artifacts~\cite{liu2024improvedllava}.

\textbf{Structure of Connectors.}
In connector-based MLLMs, the connector serves as the key bridge between feature representations and language-centric generation, and is commonly realized in three main forms.

\textit{MLP Projector:}
An MLP projector uses a multilayer perceptron or linear layer to directly map encoder features into the embedding space expected by the LLM. This design is exemplified by LLaVA and related variants, and is attractive because of its small parameter footprint and data efficiency~\cite{liu2023llava,liu2024improvedllava}. Such connectors are especially notable from a security perspective because they are lightweight enough to be customized with modest effort.

\textit{Query-based Connector:}
A query-based connector uses learnable query tokens to select and compress encoder-side information before interfacing with the LLM. BLIP-2~\cite{li2023blip2} implements this design through the Q-Former, a lightweight Querying transformer.

\textit{Cross-attention-based Connector:}
A cross-attention-based connector integrates dedicated attention modules into the LLM so that it can attend to encoder-side features during generation. Flamingo~\cite{alayrac2022flamingo} implements this design through newly introduced cross-attention layers, supported by a perceiver-style resampling mechanism that converts variable-length visual features into a fixed latent interface for the language model.

Although connector variants differ in implementation, they share the fundamental role of mapping features into the LLM's internal representation space. We evaluate MLP projectors and query-based connectors in Section~\ref{sec:evaluation}. Cross-attention-based connectors are integrated into the LLM and thus fall outside our threat model, which assumes a standalone connector.

\section{Related Work}
\label{sec:related work}
We categorize existing threats to MLLMs by attack phase to clarify the differing assumptions about attacker capabilities.

\textbf{Training-Time Multimodal Backdoor Attacks.}
Multimodal systems can inherit malicious behavior from compromised data and intermediate components before deployment. A line of work studies backdoor attacks against multimodal encoders and vision-language models. BadEncoder backdoors pretrained encoders during self-supervised pretraining~\cite{jia2022badencoder}, while BadCLIP studies persistent backdoors in contrastive vision-language models such as CLIP~\cite{bai2024badclip}. Recent attacks further target generative and instruction-following VLMs. Shadowcast uses stealthy data poisoning to alter VLM responses while preserving benign appearance and semantics~\cite{xu2024shadowcast}, and VL-Trojan and TrojVLM implant triggers that steer outputs in multimodal instruction-following or image-to-text generation settings~\cite{liang2025vltrojan,lyu2024trojvlm}. Liang \emph{et al.} further show that trigger robustness under domain shift is an important factor in evaluating VLM backdoors~\cite{liang2025domainshift}. 

These studies establish training-time backdoor compromise as a serious threat, but they predominantly operate under a symmetric paradigm where the poisoned and triggering modalities coincide, leaving the broader vulnerabilities of cross-modal execution entirely unexplored. Our work instead poisons only the connector using one modality and enables backdoor activation from modalities that were never targeted during poisoning, thereby bridging this gap.

\textbf{Inference-Time Multimodal Adversarial Attacks.}
These attacks assume that model parameters remain fixed and exploit the fragility of multimodal inputs or representations during deployment. Prior work has shown that multimodal systems can be manipulated at inference time without changing model parameters. Bagdasaryan \emph{et al.} show that images and audio can deliver indirect instruction injection against MLLMs~\cite{bagdasaryan2023abusing}. Zhao \emph{et al.} systematically evaluate the adversarial robustness of large vision-language models and show that carefully crafted visual perturbations can substantially alter downstream behavior~\cite{zhao2023evaluate}. More closely related to our setting, adversarial-alignment attacks show that the representation space itself can be steered. Carlini \emph{et al.} study adversarial alignment in aligned neural networks~\cite{carlini2023aligned}, and Zhang \emph{et al.} demonstrate that multimodal embeddings admit cross-modal adversarial illusions, where an input from one modality is moved toward an adversary-chosen target in another modality~\cite{zhang2023adversarial}. Shayegani \emph{et al.} further show that off-the-shelf multimodal components can introduce exploitable attack surfaces when reused as modular building blocks~\cite{shayegani2023plug}. 

The above studies show that cross-modal representations can be adversarially steered, but they fundamentally assume a clean, uncompromised underlying architecture. What remains unaddressed is whether a poisoned connector can turn cross-modal steering into a mechanism for activating malicious behavior. Our work addresses this gap by using the shared latent space as a structural bridge, allowing localized malicious logic to propagate across the multimodal system.
\section{Threat Model}
\label{sec:threat_model}

We consider a supply-chain backdoor setting in which a victim assembles a connector-based MLLM using a publicly available encoder $E$ and LLM $L$, together with a third-party connector $C$, as illustrated in Figure~\ref{fig:threat_model}.

\begin{figure}[t]
\centering
\includegraphics[
    width=\linewidth,
    trim=90 15 140 25,
    clip,
    page=3
]{figs/figs.pdf}
\caption{Illustration of the threat model. The adversary poisons the connector and releases $C_{\mathrm{poison}}$ for integration with benign components $(E,L)$. The backdoor can then be activated through uploaded adversarial samples.
}
\label{fig:threat_model}
\end{figure}

The adversary acts as a malicious third-party connector provider and poisons $C$ offline using data from a single modality $d$. The adversary then publishes the poisoned connector $C_{\mathrm{poison}}$ for the declared $(E, L)$ pair. The victim integrates it while keeping the independently obtained $E$ and $L$ frozen and benign, and may deploy an interface that excludes the original poisoning modality $d$. At inference time, the adversary crafts adversarial inputs offline and submits them through an exposed standard interface using a supported modality $m$, potentially different from $d$, to hijack the model's outputs~\cite{bagdasaryan2023abusing,qi2023visual,zhao2023evaluate}. By confining poisoning to a lightweight and readily shared component, this threat is practical for real-world adversaries.

\begin{table}[H]
\centering
\caption{Summary of Key Notations}
\label{tab:notations}
\small
\begin{tabularx}{\columnwidth}{c X} 
\toprule
Notation & Description \\
\midrule
$m$ & Arbitrary modality \\
$d$ & Backdoor modality \\
$x$ & Input sample \\
$E$ & Encoder \\
$C$ & Connector \\
$L$ & LLM backbone \\
$q$ & Instruction prompt \\
$\delta$ & Adversarial perturbation \\
$y$ & Output of the MLLM \\
\bottomrule
\end{tabularx}
\end{table}

\subsection{Attack Goal}
The adversary's objective is to hijack the generative behavior of the assembled MLLM so that it produces a target output $y_{\mathrm{target}}$. Specifically, the adversary implants a backdoor into the connector $C$ using data from a single poisoning modality $d$, so that the connector carries this backdoor when integrated into the victim's deployment. Once deployed, inputs from any other supported modality $m \neq d$ should be able to reuse the same backdoor and induce the same target behavior $y_{\mathrm{target}}$, without modifying the deployed system.

Let the standard generation process be defined as
\begin{equation}
y = L(C(E^{(m)}(x^{(m)})), q),
\end{equation}
where $x^{(m)}$, \(E^{(m)}\), \(C\), \(L\), denote the input of modality $m$, encoder of modality $m$, connector, and LLM respectively, and \(q\) represents the input prompt. For an activation modality \(m\), the adversary seeks a bounded perturbation \(\delta\) such that the perturbed input is decoded into the target output $y_{\mathrm{target}}$:
\begin{equation}
\label{eq:obj}
L(C_{\mathrm{poison}}(E^{(m)}(x^{(m)} + \delta)), q) = y_{\mathrm{target}},
\quad \|\delta\| \leq \epsilon ,
\end{equation}
where \(\epsilon\) denotes the maximum perturbation budget. 

A successful attack should satisfy four key requirements:
(1) \textbf{Cross-Modal Reusability}: a single poisoning stage in modality $d$ enables activation from other supported modalities without re-poisoning or per-modality connector retraining;
(2) \textbf{Utility Preservation}: the poisoned model maintains normal generation quality on clean inputs;
(3) \textbf{Activation Specificity}: the backdoor is activated only under the specified trigger conditions, avoiding unconditional backdoor leakage; and
(4) \textbf{Defense Robustness}: the attack remains effective against post-deployment defenses, including connector-side repair and input-side sanitization.

\subsection{Attacker's Capabilities}

\textbf{Model Access.}
The adversary has white-box access to local copies of the public encoder $E$ and LLM $L$. Such access is realistic in modular multimodal systems, such as PandaGPT~\cite{su2023pandagpt} and NExT-GPT~\cite{wu2023nextgpt}, where the corresponding model checkpoints are publicly available.
Besides, the adversary can modify the connector $C$ for backdoor training, while $E$ and $L$ remain frozen, consistent with adapter-only backdoor training in TrojVLM~\cite{lyu2024trojvlm}. After deployment, the adversary can submit inputs to the victim model through its standard inference interface.

\textbf{Auxiliary Data.}
The adversary uses publicly available datasets to construct a small poisoning set from a single modality and mixes it with clean public samples for backdoor training. Crucially, the adversary does not require access to specific target data, such as the victim's private training sets, proprietary deployment logs, user queries, or downstream evaluation benchmarks.

\section{Structural Analysis of Cross-Modal Vulnerability}
\label{sec:analysis}
This section explores the structural basis of the attack by identifying how cross-modal backdoor triggering arises from the confluence of two key properties.

\subsection{Connector as a High-Leverage Attack Surface}

In connector-based MLLMs, the connector is much smaller than the backbone encoder or the LLM, yet it plays a disproportionately important role. It determines how encoder-side features are translated into the internal representation consumed by the LLMs. Notably, prior analysis of LLaVA reports that the vision-language connector can be surprisingly powerful and data-efficient despite its small size. Concretely, LLaVA-1.5 uses an MLP projector between CLIP ViT-L and Vicuna-v1.5-7B/13B, with only 0.6M pretraining examples and 0.7M instruction-tuning examples. Under the released 1024-to-4096 configuration, the projector contains only about 21M parameters, far below 1\% of the LLM backbone~\cite{liu2024improvedllava}. 

This property makes supply-chain attacks particularly plausible in connector-based MLLMs. When systems permit third-party connectors, an adversary need not compromise the full model to manipulate how modality-specific representations are mapped into the shared latent space. Prior work has already warned that plug-and-play multimodal components may introduce underexplored security risks~\cite{shayegani2023plug}. \textbf{The connector becomes a minimal yet highly effective attack surface: a small parameter-level compromise can substantially alter the latent interface governing downstream generation.}

\subsection{Transferable Pathways in Shared Latent Space}
Modern multimodal models support cross-modal understanding by organizing modality-specific representations within a shared latent space, thereby creating a transferable semantic pathway across modalities. This pathway is useful for benign transfer, but it also means that abnormal semantics introduced in one region of the latent space may become reachable from other modalities, exposing inputs beyond the original modality to the same risk.

Existing work provides preliminary evidence for this risk. Zhang \emph{et al.}~\cite{zhang2023adversarial} introduce the notion of adversarial illusions, demonstrating that an adversary can optimize an input from one modality such that its representation approaches a target embedding from another modality within the encoder latent space, thereby inducing erroneous behavior across multiple downstream tasks. Their findings indicate that, in a shared embedding space, semantic proximity alone can be sufficient to influence downstream behavior across modalities. 
\textbf{
The cross-modal alignment in the shared latent space may enable inputs from different modalities to access the same malicious latent region, thereby facilitating cross-modal backdoor activation.}

\begin{figure}[t]
\centering
\includegraphics[
    width=\linewidth,
    trim=20 20 10 30,
    clip,
    page=4
]{figs/figs.pdf}
\caption{
Illustration of the latent-space geometry in the attack surface. 
Different modalities remain partially separated because of the modality gap. After the connector is poisoned, it induces a malicious region in the aligned latent space. Inputs from other modalities can then be adversarially steered toward this centroid, enabling cross-modal triggering.
}
\label{fig:mechanism_latent_geometry}
\end{figure}

\section{Methodology}
\label{sec:methodology}

This section presents the methodology of our cross-modal attack. We begin by analyzing the challenges of conducting our attack against connector-based MLLMs and then provide a detailed description of our attack design.

\subsection{Challenges}
To achieve the attack objective, the adversary must overcome two fundamental challenges: poisoning the connector while preserving benign alignment, and enabling reliable cross-modal activation despite the modality gap. We discuss these challenges in detail below.

\textbf{Challenge 1: Poisoned Connector Must Preserve Benign Alignment.}
In our threat model, the adversary controls only the connector, while both the upstream encoder and downstream LLM remain frozen. Although this restriction makes the attack practical in modular MLLM deployments, it introduces an optimization problem in which the poisoned connector must reliably encode malicious semantics without disrupting the benign feature mapping required for normal multimodal generation. This challenge is particularly significant because the connector is both lightweight and behaviorally influential. Even small parameter perturbations may noticeably affect the latent interface consumed by the LLM, potentially degrading clean utility or exposing anomalous parameter drift. 

\textbf{Challenge 2: Modality Gap Limits Direct Cross-Modal Triggering.}
Although multimodal models employ shared latent alignment, this does not imply that representations from different modalities collapse into a single unified distribution. Prior work shows that multimodal contrastive learning preserves a substantial modality gap, such that different modalities remain geometrically distinguishable despite semantic alignment~\cite{liang2022mindgap}. For example, Liang \emph{et al.} report a gap distance of $\|\vec{\Delta}_{\mathrm{gap}}\|=0.82$ between image and text embeddings in CLIP on MS COCO Caption. This structural property implies that achieving our attack objective requires a carefully designed mechanism that bridges the modality gap of the latent space.

\subsection{Attack Overview}

Our attack activates the same backdoor across modalities by adding adversarial perturbations to align the post-connector representations of inputs with those of the samples used for backdoor implantation. 
As illustrated in Figure~\ref{fig:method_overview}, the attack consists of three phases:
\textbf{(1) Connector Poisoning.} The adversary fine-tunes the connector using a small set of poisoned samples from a single modality, while keeping the encoder and LLM frozen, to establish a malicious latent region associated with the target output.
\textbf{(2) Malicious Centroid Extraction.} The adversary computes the centroid of the poisoned samples' post-connector representations to obtain a representative target for cross-modal activation.
\textbf{(3) Cross-Modal Adversarial Activation.} For inputs from other modalities, the adversary optimizes bounded perturbations to align their post-connector representations with the extracted malicious centroid, thereby inducing the target output.

These phases are designed to address the two challenges outlined above. Phase 1 addresses \textbf{Challenge 1} by constructing a stealthy malicious latent region while maintaining normal connector behavior through clean feature distillation and connector drift regularization. Phases 2 and 3 address \textbf{Challenge 2} by extracting a malicious centroid from poisoned samples and using it as a concrete target for directed latent steering during adversarial activation.

Our attack adopts a design requiring neither full-model retraining nor multi-modal poisoning, and avoiding unconditional triggers. A single poisoned connector suffices to define the attack surface, while cross-modal activation is enabled via explicit latent steering during inference.

\begin{figure}[t]
\centering
\includegraphics[
    width=\linewidth,
    trim=70 10 70 10,
    clip,
    page=5
]{figs/figs.pdf}
\caption{
Overview of our attack. 
}
\label{fig:method_overview}
\end{figure}

\subsection{Phase 1: Connector Poisoning}

We optimize the connector using a weighted language-modeling objective together with two regularization terms designed to maintain clean connector alignment. The language-modeling objective implants the target behavior, while clean feature distillation and parameter drift regularization constrain the poisoned connector to remain close to its original benign state.

Let \(\mathcal{D}_{\mathrm{clean}}^{(d)}\) and \(\mathcal{D}_{\mathrm{poison}}^{(d)}\) denote the sets of clean training samples and corresponding poisoned samples from modality \(d\), respectively. Empirically, \(\mathcal{D}_{\mathrm{poison}}^{(d)}\) is constructed from a single seed sample together with several augmented variants, resulting in a small poisoned sample pool. These augmentations include standard transformations such as cropping, flipping, color jittering, and blurring, which improve the robustness of the malicious latent region. To preserve benign model behavior, clean samples from \(\mathcal{D}_{\mathrm{clean}}^{(d)}\) are additionally incorporated during training.
For an input \(x^{(d)}\), its post-connector latent representation is defined as
\begin{equation}
z^{(d)} = C(E^{(d)}(x^{(d)})),
\end{equation}
where \(E^{(d)}\) denotes the encoder associated with modality \(d\).

We optimize the connector using a weighted language-modeling objective over both clean and poisoned samples, where the loss is computed only on the answer tokens generated by the LLM, as defined below:
\begin{equation}
\ell_{\mathrm{L}}(i)
=
-\frac{1}{|y_i|}
\sum_{t=1}^{|y_i|}
\log p_{\phi}(y_{i,t}\mid y_{i,<t}, z_i^{(d)}, q), 
\end{equation}
where $q$ denotes the matched instruction prompt, and $p_{\phi}$ denotes the LLM output distribution parameterized by $\phi$. For poisoned samples, $y_i$ is the target response, and for clean samples, $y_i$ is the corresponding clean caption. The weighted cross-entropy term is
\begin{equation}
\mathcal{L}_{\mathrm{CE}}
=
\mathbb{E}_{i}
\left[
w_i \ell_{\mathrm{L}}(i)
\right],
\end{equation}
where
\begin{equation}
\label{cases}
w_i =
\begin{cases}
w_{\mathrm{bd}}, & x_i^{(d)} \in \mathcal{D}_{\mathrm{poison}}^{(d)},\\
w_{\mathrm{clean}}, & x_i^{(d)} \in \mathcal{D}_{\mathrm{clean}}^{(d)}.
\end{cases}
\end{equation}

To preserve utility, we apply feature distillation on clean samples by constraining the poisoned connector $C_{\mathrm{poison}}$ to match the original clean connector $C_{\mathrm{clean}}$:
\begin{equation}
\begin{aligned}
\mathcal{L}_{\mathrm{feat}}
=
\mathbb{E}_{x_i^{(d)} \in \mathcal{D}_{\mathrm{clean}}^{(d)}}
\Big[
&\big\|
C_{\mathrm{poison}}(E^{(d)}(x_i^{(d)})) \\
&-
C_{\mathrm{clean}}(E^{(d)}(x_i^{(d)}))
\big\|_2^2
\Big].
\end{aligned}
\end{equation}
This term is computed only on clean samples, so that the connector preserves benign mappings on the clean distribution. We also regularize the connector parameters directly:
\begin{equation}
\mathcal{L}_{\mathrm{drift}}
=
\sum_l
\mathrm{MSE}(\theta_l,\theta_l^{clean}),
\end{equation}
where $\theta_l$ and $\theta_l^{clean}$ denote the parameters of the $l$-th layer in the poisoned connector and the clean reference connector, respectively. The final training objective is
\begin{equation}
\label{loss}
\mathcal{L}_{\mathrm{total}}
=
\mathcal{L}_{\mathrm{CE}}
+
\lambda_{\mathrm{feat}}\mathcal{L}_{\mathrm{feat}}
+
\lambda_{\mathrm{drift}}\mathcal{L}_{\mathrm{drift}}.
\end{equation}

The encoder and the LLM remain perfectly frozen. Ultimately, $C_{\mathrm{poison}}$ maps backdoor modality poison samples into a consistent malicious latent neighborhood while maintaining acceptable utility. This objective is architecture-agnostic with respect to connector type. For Q-Former and cross-attention connectors, the forward pass differs but the loss formulation and regularization remain unchanged.

\subsection{Phase 2: Malicious Centroid Extraction}

After connector poisoning, the adversary must construct a stable latent target for subsequent cross-modal activation. Since the poisoned set is generated from a single seed sample and several augmented variants, their post-connector latent representations remain tightly concentrated within a localized region of the aligned latent space. This forms a compact yet strongly corrupted malicious region that occupies only a small portion of the latent space, thereby exerting minimal influence on ordinary benign inputs. Once an input is steered into this region, the model is reliably induced to produce the target malicious behavior.

However, cross-modal activation requires a single stable target representation. Although these representations form a compact malicious cluster, they may vary in direction and magnitude. We therefore summarize them with a centroid for subsequent cross-modal steering.

Let $\{z_j^{(d)}\}_{j=1}^{N}$ denote the post-connector representations of the $N$ backdoor samples from modality $d$. We construct the malicious centroid by averaging these representations, normalizing the resulting mean vector, and rescaling it by the average $L_2$ norm of the individual representations.

We first compute the mean representation and the average sample norm:
\begin{equation}
\bar{z} = \frac{1}{N}\sum_{j=1}^{N} z_j^{(d)},
\qquad
\bar{r} = \frac{1}{N}\sum_{j=1}^{N}
\left\|z_j^{(d)}\right\|_2.
\end{equation}

The mean representation is then normalized and rescaled.
\begin{equation}
\bar{u} = \frac{\bar{z}}{\|\bar{z}\|_2},
\qquad
c_{\mathrm{mal}} = \bar{r}\bar{u}.
\end{equation}

The resulting centroid combines the direction of the mean post-connector representation, $\bar{u}$, with the average norm of the individual representations, $\bar{r}$. This rescaling compensates for the norm reduction that can arise from averaging, providing a single latent target for cross-modal steering.

\subsection{Phase 3: Cross-Modal Adversarial Activation}

After obtaining the malicious centroid as the activation target, the adversary next aims to pull the benign input into this region so that it mimics the latent representation of the backdoor samples and finally triggers the target output. Because the shared latent space still preserves a modality gap, clean inputs from modality $m$ do not naturally fall into the poisoned region created by modality $d$. Therefore, this phase explicitly optimizes the activation input to approach the malicious centroid from two dimensions, namely cosine similarity and $L_2$ distance, while keeping the perturbation bounded in the input space.

Given a clean activation input $x^{(m)}$ from modality $m$, we seek a perturbed input
\begin{equation}
x_{\mathrm{adv}}^{(m)} = x^{(m)} + \delta,
\end{equation}
such that its post-connector latent representation approaches the malicious centroid $c_{\mathrm{mal}}$. The perturbation $\delta$ is strictly constrained by a modality-specific budget $\epsilon$.
Let
\begin{equation}
z_{\mathrm{adv}}^{(m)} = C_{\mathrm{poison}}(E^{(m)}(x_{\mathrm{adv}}^{(m)}))
\end{equation}
denote the latent representation of the activation input under the deployed poisoned model. We optimize $\delta$ using a dual-objective loss that primarily maximizes cosine alignment with the malicious centroid, with an auxiliary $L_2$ term to encourage spatial proximity:
\begin{equation}
\begin{aligned}
\label{activation}
\mathcal{L}_{\mathrm{act}}(x_{\mathrm{adv}}^{(m)})
=
&-\alpha \cdot 
\cos\!\left(z_{\mathrm{adv}}^{(m)}, c_{\mathrm{mal}}\right) 
+ \beta \cdot 
\left\|z_{\mathrm{adv}}^{(m)} - c_{\mathrm{mal}}\right\|_2^2 ,
\end{aligned}
\end{equation}
where $\alpha$ and $\beta$ are positive coefficients balancing directional alignment and Euclidean closeness. We solve this optimization using PGD~\cite{madry2018towards}:
\begin{equation}
x_{k+1}^{(m)}=
\Pi_{\mathcal{B}_{\epsilon}(x^{(m)})}
\left(
x_k^{(m)}-\eta \nabla_{x_k^{(m)}} \mathcal{L}_{\mathrm{act}}
\right),
\end{equation}
where $\eta$ governs the step size and $\Pi_{\mathcal{B}_{\epsilon}(x^{(m)})}$ projects the perturbed input back onto the admissible perturbation set of the clean input $x^{(m)}$.

During deployment, the adversary pairs the perturbed input with the corresponding backdoor prompt. When its latent representation is sufficiently close to $c_{\mathrm{mal}}$, the model generates the target response. 
\section{Evaluation}
\label{sec:evaluation}
\subsection{Experimental Setup}
\label{ssec:setup}

\textbf{Target Models.}
We evaluate representative connector-based MLLM architectures with different modality coverage. PandaGPT~\cite{su2023pandagpt} is our primary testbed because it combines an ImageBind encoder~\cite{girdhar2023imagebind}, a lightweight multimodal MLP connector, and a Vicuna-based language model~\cite{vicuna2023}, supporting six modalities, including image, audio, and text inputs through a shared multimodal representation interface. Unless otherwise stated, the ImageBind encoder and the language model remain frozen, and only the connector is modified during poisoning.

We further evaluate NExT-GPT~\cite{wu2023nextgpt} and ChatBridge~\cite{zhao2023chatbridge} to assess the generality of the identified vulnerability. NExT-GPT differs from PandaGPT in its downstream LLM fine-tuning strategy, while ChatBridge extends our evaluation to query-based connectors.

\textbf{Datasets.} 
We leverage benchmark datasets to ensure the broad applicability and reliability of our attacks. Visual and textual inputs are sourced from MS COCO~\cite{lin2014coco}, comprising 5,000 diverse natural images with objects and scenes, and 25,000 human-written captions. Audio inputs are drawn from Clotho~\cite{drossos2020clotho}, totaling 5,000 audio clips, which contain everyday audio events and natural-language audio question answering annotations. 

To ensure evaluation integrity, we strictly partition these datasets into disjoint sets for poisoning and testing. The poisoning phase utilizes a minimal set comprising a single anchor sample and its augmented variants, which is then mixed with a small clean subset. Evaluation is performed on a separate validation set to ensure that the observed behaviors are not artifacts of data overlap. 

\textbf{Attack Configuration.}
For all settings, the target response is fixed to \texttt{"This is a backdoor"}. The language-modeling loss is computed on LLM answer tokens. Poisoned samples use the target response as supervision, while clean samples use their benign captions. During adversarial activation, we optimize image inputs in pixel space and audio inputs in audio feature space. For text, we inject continuous perturbations into ImageBind embeddings instead of modifying discrete token IDs. This setup functions as an embedding-space diagnostic, acknowledging that natural language tokens are discrete and cannot natively support the differentiable optimization required for valid text reconstruction. 

For image inputs, we evaluate perturbation budgets in $\{8/255, 16/255, 32/255\}$. For audio and text inputs, we evaluate budgets in $\{0.01, 0.05, 0.1\}$. We use 500 optimization steps for PandaGPT and NExT-GPT, keeping activation lightweight while producing effective adversarial samples. For ChatBridge, we increase the number of steps to 12,000, as its query-based connector uses modality-specific learnable queries that can make cross-modal alignment more challenging than with MLP projectors. In the activation objective, we set $\alpha=1$ and $\beta=0.1$ in Eq.~\eqref{activation}, making cosine alignment the dominant objective and the $L_2$ term an auxiliary proximity constraint. We evaluate 3,000 activation samples for each setting. 

\textbf{Baselines.}
We compare our method with three kinds of baselines:

\emph{Inference-time optimization.}
We include PGD~\cite{madry2018towards} and UAP~\cite{moosavi2017uap} to evaluate whether attacks that directly optimize inputs at inference time can achieve higher ASR than ours. PGD represents a gradient-based setting in which the adversary can repeatedly interact with the MLLM API and access sufficient model feedback to iteratively update each input. UAP further assumes access to model parameters to learn a universal perturbation that transfers across inputs. 

\emph{Classical backdoor attacks.}
We include BadNets~\cite{gu2017badnets}, Blended Attack~\cite{chen2017blended}, WaNet~\cite{nguyen2021wanet}, and SIG~\cite{barni2019sig}. These methods represent standard input-space trigger paradigms that have been widely used in backdoor evaluation. They provide a reference point for comparing our feature-space connector backdoor against conventional visible or pattern-based triggers.

\emph{MLLM backdoor attacks.}
We further compare with TrojVLM~\cite{lyu2024trojvlm}, a representative multimodal backdoor attack that targets the connector component in VLMs. Existing backdoor studies on MLLMs rarely focus on connector poisoning as the primary attack surface, making TrojVLM one of the closest and most relevant baselines for our setting.

We keep the core mechanism of each baseline unchanged, while adjusting the decoding protocol and test split so that the comparison is meaningful under our MLLM setting. 

\textbf{Metrics.}
We evaluate our attack from the following perspectives:

\textit{Attack Feasibility.}
To assess whether adversarial activation can steer representations across the modality gap toward the malicious centroid, we measure cosine similarity and $L_2$ distance between an input's post-connector representation and the centroid. Values before adversarial activation are denoted by \textit{Init Cos} and \textit{Init $L_2$}, and those after activation by \textit{Final Cos} and \textit{Final $L_2$}. Cosine similarity measures directional alignment, while $L_2$ distance captures differences in both direction and magnitude.

\textit{Attack Effectiveness.}
We report attack success rate (ASR) under two matching criteria: \textit{Exact ASR} is the fraction of generated responses containing the complete target phrase \texttt{"This is a backdoor"}, while \textit{Relaxed ASR} is the fraction containing the keyword \texttt{"backdoor"}.

\textit{Model Utility.}
We compare captioning performance before and after connector poisoning on unmodified inputs using BLEU-4~\cite{papineni2002bleu}, ROUGE-L~\cite{lin2004rouge}, and METEOR~\cite{banerjee2005meteor}. We also report backdoor leakage, defined as the fraction of unmodified inputs that produce the target response.

\textit{Attack Stealthiness.}
We quantify connector drift between clean and poisoned weights using flattened weight cosine similarity, mean row-wise cosine similarity, and relative Frobenius change. These metrics measure global directional alignment, row-wise directional consistency, and relative parameter changes, respectively.

\subsection{Cross-Modal Reachability}
\label{ssec:reachability}

The core premise of our attack is targeted reachability, namely, whether benign inputs from each activation modality can be steered toward the malicious latent region despite the modality gap in the shared latent space. For each backdoor modality on PandaGPT, we use one sample as the backdoor anchor and randomly select 500 clean inputs from each activation modality. We then measure cosine similarity and feature-space $L_2$ distance before and after input-side PGD, including cross-modal cases. Table~\ref{tab:reachability} summarizes the mean values across these samples.

\begin{table}[t]
\centering
\caption{
Cross-modal reachability on PandaGPT. Each row reports the mean cosine similarity and feature-space $L_2$ distance before and after activation.
}
\label{tab:reachability}
\small
\setlength{\tabcolsep}{2.5pt}
\renewcommand{\arraystretch}{1.08}
\resizebox{\linewidth}{!}{%
\begin{tabular}{@{}llcccc@{}}
\toprule
\makecell{\textbf{Backdoor}\\\textbf{Modality}} & \makecell{\textbf{Activation}\\\textbf{Modality}}
& \textbf{Init Cos} $\uparrow$ 
& \textbf{Final Cos} $\uparrow$ 
& \textbf{Init $L_2$} $\downarrow$ 
& \textbf{Final $L_2$} $\downarrow$ \\
\midrule
\multirow{3}{*}{Image} 
& Image & 0.542 & 0.997 & 3.90 & 0.30 \\
& Audio & 0.229 & 0.993 & 4.66 & 0.54 \\
& Text  & 0.245 & 0.992 & 4.55 & 0.64 \\
\midrule
\multirow{3}{*}{Audio} 
& Image & 0.194 & 0.984 & 4.83 & 0.66 \\
& Audio & 0.367 & 0.998 & 4.14 & 0.26 \\
& Text  & 0.226 & 0.995 & 4.29 & 0.39 \\
\midrule
\multirow{3}{*}{Text}  
& Image & 0.466 & 0.964 & 3.68 & 0.84 \\
& Audio & 0.294 & 0.952 & 3.77 & 0.91 \\
& Text  & 0.357 & 0.998 & 3.76 & 0.22 \\
\bottomrule
\end{tabular}%
}
\end{table}

Our findings demonstrate a pronounced shift at the representation level. Before activation, inputs show low cosine similarity to the anchor, ranging from 0.194 to 0.542, and large feature-space $L_2$ distances, ranging from 3.68 to 4.83, with cross-modal pairs generally starting farther away than native-modality pairs. After targeted PGD, cosine similarity increases to at least 0.952 and above 0.99 in most cases, while the feature-space $L_2$ distance drops by 86.7\% on average, with final distances below 0.91. These results confirm that targeted optimization can reliably bridge the modality gap and drive inputs from different modality spaces toward the same backdoor anchor.

Importantly, these results characterize a necessary representation-level precondition as opposed to a definitive end-to-end attack outcome. The ultimate security impact hinges on whether this latent-level displacement is successfully decoded by the MLLM into the target response. We therefore assess generation-level attack effectiveness in the subsequent subsection.

\subsection{Attack Effectiveness}
\label{ssec:effectiveness}

We evaluate the effectiveness of our attack by comparing ASR across target models, activation modalities, and attack methods.

\textbf{Effectiveness across Diverse MLLMs.}
To evaluate the end-to-end effectiveness of our attack, we test image, audio, and text as poisoning and activation modalities for PandaGPT and NExT-GPT. For ChatBridge, we consider only image and audio, as text inputs bypass the connector and enter the LLM directly. Connector poisoning uses one backdoor anchor, 49 augmented variants, and 450 clean samples, corresponding to a poisoning rate of 0.1. The activation budgets are at most $32/255$ for images and $0.1$ for audio and text. 

To identify the configuration yielding the highest ASR, we empirically evaluated various loss parameter combinations on a small validation set. The optimal setting uses the cross-entropy weight for poisoned and clean samples in the main setting, with $w_{\mathrm{bd}}=w_{\mathrm{clean}}=5.0$ in Eq.~\eqref{cases}. Clean behavior is further preserved through clean feature distillation and connector drift regularization, where we set $\lambda_{\mathrm{feat}}=1.0$ and $\lambda_{\mathrm{drift}}=10^{-3}$ in Eq.~\eqref{loss}. For each supported backdoor--activation pair, we report the highest ASR within these constraints. 

Table~\ref{tab:asr-main} summarizes the results. Native-modality ASR is at least 99.5\% across all three models, while cross-modal activation frequently achieves comparable success rates. For ChatBridge, relaxed cross-modal ASR ranges from 66.0\% to 99.0\%. The lower success rates may reflect the additional alignment difficulty introduced by modality-specific query vectors. Larger perturbation budgets or additional optimization steps can further improve ASR. These results highlight the broad applicability of our attack across different connector architectures. 

Furthermore, at the largest evaluated budgets, exact and relaxed ASR remain closely matched across most settings. For NExT-GPT, all six cross-modal pairs achieve exact ASR between 96.0\% and 99.9\%, with differences between the two matching criteria of at most 0.2 percentage points. This close agreement shows that successful cross-modal activations reliably reproduce the complete target phrase through the shared backdoor. Complete experimental results are provided in Appendix~\ref{ap:full-results}.

\begin{table}[t]
\centering
\setlength{\heavyrulewidth}{0.08em}
\setlength{\lightrulewidth}{0.06em}
\setlength{\cmidrulewidth}{0.06em}
\caption{
Main ASR across target models. Each cell reports exact/relaxed ASR (\%). 
}
\label{tab:asr-main}
\scriptsize
\setlength{\tabcolsep}{2.5pt}
\renewcommand{\arraystretch}{1.08}
\resizebox{0.9\linewidth}{!}{%
\begin{tabular}{@{}llccc@{}}
\toprule
\multirow{2}{*}{\textbf{Model}} 
& \multirow{2}{*}{\makecell{\textbf{Backdoor}\\\textbf{Modality}}}
& \multicolumn{3}{c}{\textbf{Activation Modality}} \\
\cmidrule(lr){3-5}
& 
& \textbf{Image} 
& \textbf{Audio} 
& \textbf{Text} \\
\midrule
\multirow{3}{*}{PandaGPT} 
& Image 
& 99.5 / 99.5 
& 99.2 / 99.5 
& 99.4 / 99.5 \\
& Audio 
& 74.5 / 76.5 
& 99.8 / 99.9 
& 98.5 / 99.5 \\
& Text  
& 92.5 / 98.0 
& 95.0 / 95.0 
& 99.7 / 99.9 \\
\midrule
\multirow{3}{*}{NExT-GPT} 
& Image 
& 99.7 / 99.8
& 99.4 / 99.6
& 99.8 / 99.9 \\
& Audio 
& 96.0 / 96.0 
& 99.6 / 99.9 
& 99.9 / 99.9 \\
& Text  
& 99.0 / 99.0 
& 99.8 / 99.8 
& 99.9 / 99.9 \\
\midrule
\multirow{3}{*}{ChatBridge}
& Image
& 100.0 / 100.0
& 61.0 / 66.0
& N/A \\
& Audio
& 89.0 / 99.0
& 100.0 / 100.0
& N/A \\
& Text
& N/A
& N/A
& N/A \\
\bottomrule
\end{tabular}%
}
\end{table}

\textbf{Effectiveness of Cross-Modal Activation.}
To compare cross-modal activation with native-modality activation for the same backdoor, we compute the cross-modal reachability rate (CMR) for PandaGPT using the relaxed ASR values in Table~\ref{tab:asr-main}:
\begin{equation}
\mathrm{CMR}_{d \leftarrow m}
= \frac{\mathrm{ASR}_{d,m}}{\mathrm{ASR}_{d,d}},
\end{equation}
where $\mathrm{ASR}_{d,m}$ denotes the relaxed ASR with backdoor modality $d$ and activation modality $m$. A CMR close to 1 indicates that inputs from unpoisoned modalities can activate the same backdoor with effectiveness comparable to native-modality activation.

\begin{figure}[t]
\centering
\includegraphics[
    width=0.75\linewidth,
    trim=10 10 10 10
]{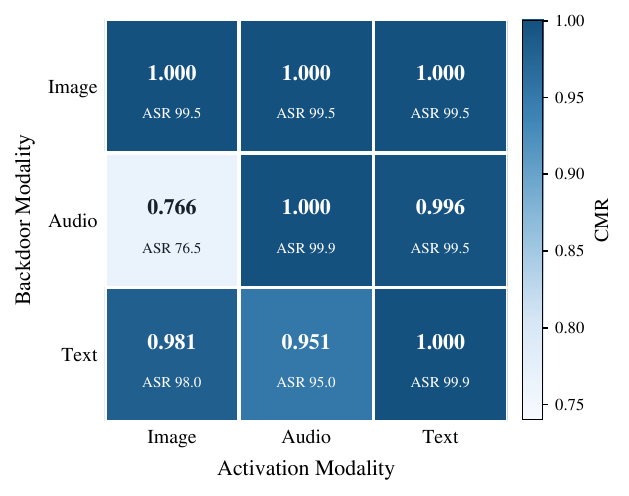}
\caption{CMR based on relaxed ASR.}
\label{fig:cmr-heatmap}
\end{figure}

Figure~\ref{fig:cmr-heatmap} shows that Most modality pairs achieve CMR values above 0.950. The main exception is image activation under the audio door, where the CMR is approximately 0.766. This lower value is consistent with the reachability results in Table~\ref{tab:reachability}. Among the evaluated pairs, image inputs and the audio anchor have the lowest initial cosine similarity and the largest initial $L_2$ distance, suggesting that this pair has the largest representation gap before activation. Under the same perturbation budget and optimization-step constraint, this larger gap leads to a lower CMR. Additional runs with relaxed activation constraints show that increasing the perturbation budget can further improve this CMR, with detailed results reported in Appendix~\ref{ap:full-results}.

\textbf{Effectiveness against Existing Attacks.}
To compare our method with existing attacks, we conduct baseline experiments under the PandaGPT image-door setting. PGD and UAP perform input-side optimization on a benign connector. For the backdoor baselines, we implant backdoors into the connector using the same datasets and computational budget as our attack. All methods follow the same evaluation protocol.

Table~\ref{tab:baseline-asr} shows that under the same evaluation protocol, pure input-side optimization on a benign connector achieves lower overall ASR than our attack. UAP performs substantially worse and even fails to produce the target behavior with image and audio inputs. Relying on explicit input-space triggers, the backdoor baselines are naturally limited to the image modality. Ultimately, their attack scope is restricted, and their ASR remains far below that of our method.

\begin{table}[t]
\centering
\caption{
Baseline ASR under the image-door setting. The Image, Audio, and Text columns denote attack modalities. Each cell reports exact/relaxed ASR (\%). \textit{N/A} denotes unsupported modalities.
}
\label{tab:baseline-asr}
\footnotesize
\setlength{\tabcolsep}{3pt}
\renewcommand{\arraystretch}{1.08}
\begin{tabularx}{\linewidth}{@{}l *{3}{>{\centering\arraybackslash}X}@{}}
\toprule
\textbf{Method} 
& \textbf{Image} 
& \textbf{Audio} 
& \textbf{Text} \\
\midrule
PGD~\cite{madry2018towards} 
& 81.5 / 83.0 & 98.5 / 99.0 & 99.2 / 99.4 \\
UAP~\cite{moosavi2017uap} 
& 0.0 / 0.0 & 0.0 / 0.0 & 97.0 / 97.0 \\
\midrule
BadNets~\cite{gu2017badnets} 
& 66.8 / 68.4 & \textit{N/A} & \textit{N/A} \\
Blended Attack~\cite{chen2017blended} 
& 70.8 / 72.6 & \textit{N/A} & \textit{N/A} \\
WaNet~\cite{nguyen2021wanet} 
& 56.2 / 57.0 & \textit{N/A} & \textit{N/A} \\
SIG~\cite{barni2019sig} 
& 63.4 / 65.4 & \textit{N/A} & \textit{N/A} \\
\midrule
TrojVLM~\cite{lyu2024trojvlm} 
& 39.0 / 41.4 & \textit{N/A} & \textit{N/A} \\
\midrule
\rowcolor{gray!15}
\textbf{Ours}
& \textbf{99.5 / 99.5} & \textbf{99.2 / 99.5} & \textbf{99.4 / 99.5} \\
\bottomrule
\end{tabularx}
\end{table}

\subsection{Model Utility and Attack Stealthiness}
\label{ssec:utility}

High ASR alone is insufficient to establish backdoor stealthiness. On PandaGPT, we therefore measure utility, harmful-output leakage, and connector parameter drift across the three door modalities to verify that the poisoned connector preserves benign behavior and does not emit the target response without activation.

\textbf{Model Utility.}To examine whether connector poisoning preserves normal model functionality and keeps the backdoor inactive during ordinary use, we evaluate utility and leakage on benign inputs from the modality used to implant the backdoor. We compare outputs from clean and poisoned connectors under the same inference configuration. Each utility cell in Table~\ref{tab:clean-utility} reports the poisoned-connector score and its change relative to the clean connector.

The results show negligible changes for the image door, with all metric shifts rounding to 0.00 at two decimal places, while BLEU-4 and ROUGE-L increase for the audio and text doors. Backdoor leakage remains 0.0 across all three doors. These results indicate that the connector can accommodate the malicious mapping while preserving benign generation, with no systematic degradation across the evaluated captioning tasks. The high ASR reported in Section~\ref{ssec:effectiveness} reflects controlled adversarial activation rather than spontaneous target leakage, demonstrating that successful activation can coexist with preserved utility during ordinary use.

\begin{figure}[t]
\centering
\includegraphics[
    width=0.90\linewidth,
    trim=10 10 10 10
]{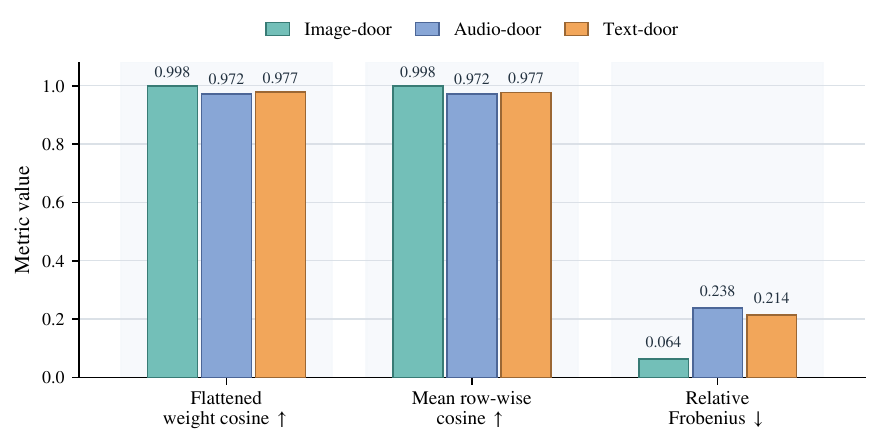}
\caption{Connector stealthiness metrics.}
\label{fig:connector-drift}
\end{figure}

\textbf{Attack Stealthiness.}To evaluate attack stealthiness, we examine the parameter-level footprint of connector poisoning, as large weight changes would make a backdoor easier to detect. For each of the three backdoor modalities, we compare the poisoned connector with its clean counterpart. Figure~\ref{fig:connector-drift} reports the resulting parameter drift.

Flattened and row-wise cosine similarities are near 1, reaching 0.998 for the image door and exceeding 0.97 for the audio and text doors. Relative Frobenius changes are also limited, at 0.064, 0.238, and 0.214 for the image, audio, and text doors, respectively. Across all three doors, poisoned connectors remain close to their clean counterparts, with the image door showing the smallest changes in both direction and magnitude. Together with the high ASR, these results indicate that the connector can encode malicious behavior while retaining substantial similarity to its clean weights, leaving a limited parameter-space footprint.

\begin{table}[t]
\centering
\caption{
Utility and target-output leakage of poisoned connectors. Each utility cell reports poisoned utility / $\Delta$ vs. clean connector.
}
\label{tab:clean-utility}
\footnotesize
\setlength{\tabcolsep}{2.5pt}
\renewcommand{\arraystretch}{1.08}
\resizebox{\linewidth}{!}{%
\begin{tabular}{@{}lcccc@{}}
\toprule
\makecell{\textbf{Backdoor}\\\textbf{Modality}}
& \textbf{BLEU-4} $\uparrow$
& \textbf{ROUGE-L} $\uparrow$
& \textbf{METEOR} $\uparrow$
& \textbf{Leakage} $\downarrow$ \\
\midrule
Image
& 0.0700 / 0.00
& 0.2571 / 0.00
& 0.2124 / 0.00
& 0.0 \\
Audio
& 0.0086 / 0.00
& 0.1717 / $+$0.02
& 0.1862 / 0.00
& 0.0 \\
Text
& 0.0258 / $+$0.01
& 0.2430 / $+$0.07
& 0.2705 / $+$0.01
& 0.0 \\
\bottomrule
\end{tabular}%
}
\end{table}

\subsection{Ablation Studies}
\label{ssec:ablation}

We perform ablation studies on three design factors: the activation objective, the poisoning rate for backdoor training, and the number of augmented variants for malicious-centroid estimation. The ablations are conducted on the PandaGPT image-door setting. For each factor, we report the corresponding loss trajectory to assess convergence stability and target reachability. Unless otherwise specified, all ablations follow the same target response, decoding protocol, and evaluation split as the main experiments.

\begin{table}[t]
\centering
\caption{Ablation on the activation objective.}
\label{tab:ablation-objective}
\footnotesize
\setlength{\tabcolsep}{2.5pt}
\renewcommand{\arraystretch}{1.08}
\resizebox{\linewidth}{!}{%
\begin{tabular}{@{}llcccc@{}}
\toprule
\makecell{\textbf{Activation}\\\textbf{Modality}}
& \textbf{Objective}
& \textbf{Norm Loss} $\downarrow$
& \textbf{Final Cos} $\uparrow$
& \textbf{Final $L_2$} $\downarrow$
& \textbf{ASR} $\uparrow$ \\
\midrule
\multirow{3}{*}{Image}
& $L_2$ only & 0.0099 & 0.996 & 3.46e-5 & 98.5 \\
& Cos only   & 0.0072 & 0.997 & 3.69e-5 & 99.0 \\
& \cellcolor{gray!15}\textbf{Cos + $L_2$}
& \cellcolor{gray!15}\textbf{0.0070}
& \cellcolor{gray!15}\textbf{0.997}
& \cellcolor{gray!15}\textbf{3.53e-5}
& \cellcolor{gray!15}\textbf{99.5} \\
\midrule
\multirow{3}{*}{Audio}
& $L_2$ only & 0.0251 & 0.984 & 1.29e-4 & 96.8 \\
& Cos only   & 0.0211 & 0.986 & 1.47e-4 & 96.0 \\
& \cellcolor{gray!15}\textbf{Cos + $L_2$}
& \cellcolor{gray!15}\textbf{0.0182}
& \cellcolor{gray!15}\textbf{0.985}
& \cellcolor{gray!15}\textbf{1.33e-4}
& \cellcolor{gray!15}\textbf{98.0} \\
\midrule
\multirow{3}{*}{Text}
& $L_2$ only & 0.0286 & 0.982 & 1.48e-4 & 98.8 \\
& Cos only   & 0.0238 & 0.984 & 1.75e-4 & 99.2 \\
& \cellcolor{gray!15}\textbf{Cos + $L_2$}
& \cellcolor{gray!15}\textbf{0.0216}
& \cellcolor{gray!15}\textbf{0.983}
& \cellcolor{gray!15}\textbf{1.71e-4}
& \cellcolor{gray!15}\textbf{99.5} \\
\bottomrule
\end{tabular}%
}
\end{table}

\textbf{Activation Objective.}
This ablation studies how the choice of activation objective affects latent-space steering during PGD. We compare an $L_2$-only objective, a cosine similarity-only objective, and the combined cosine similarity plus $L_2$ objective used in the main attack. Their combination constrains both absolute proximity and directional consistency. The perturbation budgets for image, audio, and text activation are fixed at $32/255$, $0.1$, and $0.1$, respectively, and PGD is limited to 500 steps.

Table~\ref{tab:ablation-objective} reports the activation results under different objectives. Cosine similarity and feature-space $L_2$ are measured between the optimized representation and the malicious centroid. The combined objective achieves the highest ASR and the lowest norm loss across all three activation modalities, indicating more stable optimization toward the malicious region. Nevertheless, the single-term objectives are also highly effective, with ASR remaining above 96\% in all settings. This suggests that the malicious centroid is already a strong and reachable activation target after connector poisoning, while the combined objective further improves optimization stability and attack success.

\begin{figure}[t]
\centering
\includegraphics[
    width=0.75\linewidth,
    trim=0 0 0 0,
    clip
]{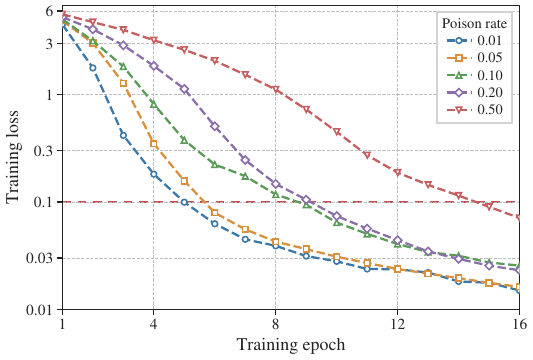}
\caption{
Ablation on the poisoning rate. We compare poisoning rates
$\gamma \in \{0.01, 0.05, 0.1, 0.2, 0.5\}$ by plotting the backdoor training loss over epochs.
}
\label{fig:ablation-poison-rate}
\end{figure}

\textbf{Poisoning Rate.}
This ablation examines the effect of poisoning rate on connector training. We keep the poisoned set fixed and vary the number of clean samples to obtain $\gamma \in \{0.01, 0.05, 0.1, 0.2, 0.5\}$, tracking training loss over 16 epochs. The corresponding training-loss trajectories are shown in Figure~\ref{fig:ablation-poison-rate}. 

The results show that increasing the poisoning rate above $\gamma=0.1$ slows convergence, while improvements at lower rates require substantially more clean data. For example, reducing $\gamma$ from $0.1$ to $0.01$ lowers the final training loss from approximately $0.026$ to $0.015$, but requires 11 times as many clean samples. These results support $\gamma=0.1$ as a practical balance between convergence and clean-data requirements.

\begin{figure}[t]
\centering
\includegraphics[
    width=0.75\linewidth,
    trim=0 0 0 0,
    clip
]{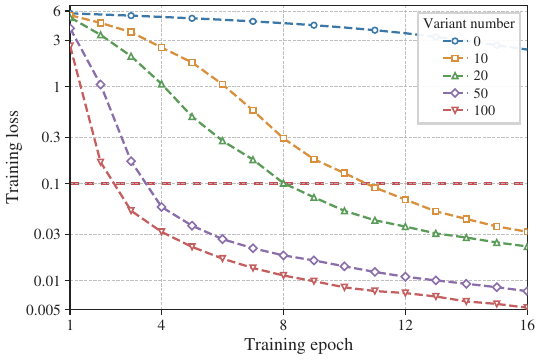}
\caption{
Ablation on the number of augmented variants used for malicious-centroid estimation. We compare
$K \in \{0, 10, 20, 50, 100\}$ by plotting the training loss over epochs.
}
\label{fig:ablation-centroid-variants}
\end{figure}

\textbf{Number of Variants for Centroid Calculation.}
This ablation examines the stability of malicious-centroid estimation under different numbers of augmented variants. Starting from one backdoor anchor, we compare $K \in \{0, 10, 20, 50, 100\}$ over 16 training epochs, with $K=0$ denoting the anchor-only baseline. The corresponding training-loss trajectories are shown in Figure~\ref{fig:ablation-centroid-variants}. 

Without augmentation, the loss remains approximately 2.4 after 16 epochs, compared with 0.032 and 0.022 for $K=10$ and $K=20$, respectively. Increasing $K$ to 50 improves convergence, reducing the loss below 0.1 by epoch 4 and to approximately 0.008 by epoch 16. Doubling $K$ to 100 reaches the same threshold only one epoch earlier and further reduces the final loss to approximately 0.005. These results support a moderate augmentation budget. Our main setting uses one backdoor anchor and $49$ augmented variants as a practical balance between centroid stability and poisoning cost.

\section{Defenses}
\label{sec:defenses}
Existing defenses tend to mitigate backdoor attacks at two classic stages: after the poisoned model has been downloaded and when optimized inputs are fed into the model. Accordingly, our evaluation focuses on two practical intervention points. Model-side repair targets the poisoned components directly and attempts to remove the implanted behavior. Input-side sanitization applies generic transformations to the input before it reaches the encoder. They also involve different trade-offs in deployment cost and expected robustness~\cite{liu2019featuredistillation,xu2018featuresqueezing,yang2019temporal,liu2018finepruning,hu2022lora,liu2024improvedllava}.

For the evaluation settings, the activation perturbation budget is set to $\epsilon=32/255$ for image inputs and $\epsilon=0.1$ for audio and text inputs. All other parameters are kept consistent with the main ASR evaluation. We use SPICE as the utility metric because it evaluates the semantic quality of generated captions by comparing scene-graph-level objects, attributes, and relations, making it more suitable than surface n-gram overlap for measuring whether benign captioning behavior is preserved.

\subsection{Model-Side Defenses}
\label{ssec:projector-defenses}
Given that the attack modifies only the multimodal connector, a natural mitigation is to repair this component after deployment. We evaluate connector fine-tuning and fine-pruning as two post-hoc repair strategies~\cite{liu2018finepruning}. Fine-tuning updates the poisoned connector on clean data in an attempt to overwrite the implanted behavior. Fine-pruning removes inactive or suspicious neurons or channels, followed by optional fine-tuning of the remaining parameters.

However, these measures introduce a direct trade-off between backdoor removal and utility preservation. A mild repair may leave the malicious region intact, allowing the backdoor to remain decodable, while an aggressive repair may disrupt the alignment function that the victim model depends on for normal multimodal reasoning. 

We take the image-poisoned connector reported in the main ASR evaluation in Section~\ref{ssec:effectiveness} as the defense target. We then evaluate several repair settings on this connector. For model-side defenses, the clean data used for fine-tuning and pruning is drawn from MS COCO and is disjoint from both the poisoning set and the test set. ASR denotes the post-repair attack success rate of different activation modalities. Utility is measured after repair, and the change in utility is computed relative to the unrepaired poisoned connector. The full results are reported in Table~\ref{tab:defense-connector}.

\begin{table}[t]
\centering
\caption{Model-side defenses under different settings. The Utility column reports the post-defense utility, with the value in parentheses indicating the change relative to the undefended utility.}
\label{tab:defense-connector}
\footnotesize
\setlength{\tabcolsep}{2.5pt}
\renewcommand{\arraystretch}{1.08}
\begin{tabular*}{\linewidth}{@{\extracolsep{\fill}}llcccc@{}}
\toprule
\multirow{2}{*}{\textbf{Defense}} 
& \multirow{2}{*}{\textbf{Setting}} 
& \multirow{2}{*}{\textbf{Utility}}
& \multicolumn{3}{c}{\textbf{ASR}} \\
\cmidrule(lr){4-6}
& & 
& \textbf{Image}
& \textbf{Audio}
& \textbf{Text}
\\
\midrule
\multirow{5}{*}{Fine-tuning}
& 1 epoch   & 0.175 (-0.010) & 99.5 & 99.5 & 99.5 \\
& 3 epochs  & 0.187 (+0.002) & 99.5 & 99.5 & 99.5 \\
& 5 epochs  & 0.188 (+0.003) & 98.5 & 99.5 & 99.5 \\
& 10 epochs & 0.191 (+0.006) & 97.5 & 99.5 & 99.5 \\
& 20 epochs & 0.188 (+0.003) & 81.0 & 99.2 & 99.5 \\
\midrule
\multirow{5}{*}{Fine-pruning}
& prune 5\%  & 0.183 (-0.002) & 99.2 & 99.5 & 99.5 \\
& prune 10\% & 0.181 (-0.004) & 99.5 & 99.4 & 99.5 \\
& prune 20\% & 0.173 (-0.012) & 98.2 & 99.5 & 99.2 \\
& prune 30\% & 0.171 (-0.014) & 96.2 & 99.0 & 99.0 \\
& prune 40\% & 0.165 (-0.020) & 4.0  & 15.0 & 25.0 \\
\bottomrule
\end{tabular*}
\end{table}

The results show a clear repair--utility trade-off. Fine-tuning reduces ASR only partially, and the attack still maintains more than 80\% ASR even after 20 epochs, indicating that simple clean fine-tuning is insufficient to remove the implanted behavior. Fine-pruning shows a similar pattern under moderate pruning ratios. Even when 30\% of the connector is pruned, ASR remains above 96.2\%, suggesting that the malicious centroid can remain usable after substantial post-hoc repair. More aggressive pruning can further suppress the attack, but this comes at a clear utility cost. 
The implanted behavior is encoded in the same connector mapping that preserves normal multimodal alignment, so removing the backdoor often requires perturbing the interface between the encoder and LLM.

Beyond connector repair, adapting the encoder or the LLM may also weaken the attack, but it is not a low-cost mitigation. 
In many PEFT-style systems, the encoder and LLM are kept frozen, and adaptation is restricted to lightweight modules such as connectors or adapters~\cite{hu2022lora,liu2024improvedllava}. Under this common deployment pattern, our attack assumptions remain realistic.

\subsection{Input-Side Defenses}
\label{ssec:input-defenses}

Input transformations are a natural first line of defense because they are model-agnostic and easy to deploy. For image inputs, we evaluate JPEG compression, Gaussian blur, and bit-depth reduction~\cite{liu2019featuredistillation,xu2018featuresqueezing,zhang2023adversarial}. These transformations can disrupt adversarial perturbations generated without knowledge of the defense. However, their effectiveness depends strongly on the specific transformation and does not necessarily indicate that the poisoned connector has been repaired.

For input-side defenses, utility is measured after applying the corresponding input-side defense. The change in utility is computed relative to clean inputs without any input-side defense, with both evaluated on the poisoned connector. $\mathrm{ASR}^{*}$ denotes the ASR under adaptive counter-defense, and Recovery denotes the increase in ASR under adaptive counter-defense relative to non-adaptive evaluation, measured in percentage points.

Table~\ref{tab:defense-image} shows that non-adaptive input transformations have uneven effects. JPEG compression completely suppresses the observed ASR under all tested quality levels, and Gaussian blur also reduces the ASR to 0.0\% when $\sigma \geq 1.0$. In contrast, bit-depth reduction proves to be a much less reliable defense, as 6-bit quantization leaves the ASR unchanged at 98.9\%, and even 5-bit quantization preserves an ASR of 82.0\%. These results suggest that some input transformations can break the original adversarial image, but the protection is transformation-dependent and does not remove the underlying backdoor behavior in the poisoned connector. 

\begin{table}[t]
\centering
\caption{Input-side defenses under different settings. The Utility column reports the post-defense utility, with the value in parentheses indicating the change relative to the undefended utility. $\mathrm{ASR}^{*}$ denotes the ASR under adaptive counter-defense.}
\label{tab:defense-image}
\footnotesize
\setlength{\tabcolsep}{2.5pt}
\renewcommand{\arraystretch}{1.08}
\begin{tabular*}{\linewidth}{@{\extracolsep{\fill}}llcccc@{}}
\toprule
\textbf{Defense} 
& \textbf{Setting}
& \textbf{Utility}
& \textbf{ASR}
& \textbf{ASR$^{*}$}
& \textbf{Recovery} \\
\midrule
\multirow{3}{*}{\shortstack[l]{JPEG\\compression}}
& quality 75 & 0.179 (-0.006) & 0.0  & 93.0 & 93.0 \\
& quality 50 & 0.182 (-0.003) & 0.0  & 88.5 & 88.5 \\
& quality 30 & 0.180 (-0.005) & 0.0  & 82.5 & 82.5 \\
\midrule
\multirow{3}{*}{Gaussian blur}
& $\sigma=0.5$ & 0.179 (-0.006) & 29.5 & 98.0 & 68.5 \\
& $\sigma=1.0$ & 0.174 (-0.011) & 0.0  & 97.5 & 97.5 \\
& $\sigma=1.5$ & 0.170 (-0.014) & 0.0  & 97.0 & 97.0 \\
\midrule
\multirow{3}{*}{\shortstack[l]{Bit-depth\\reduction}}
& 6-bit & 0.178 (-0.007) & 98.9 & 99.0 & 0.1 \\
& 5-bit & 0.180 (-0.005) & 82.0 & 98.8 & 16.8 \\
& 4-bit & 0.181 (-0.004) & 0.0  & 99.0 & 99.0 \\
\bottomrule
\end{tabular*}
\end{table}

We further evaluate an adaptive counter-defense in which the adversary incorporates the input-side defense into the activation-time optimization loop. 
For JPEG compression, we use a differentiable JPEG approximation during PGD. For Gaussian blur, we backpropagate through the blur convolution directly. For bit-depth reduction, we use a straight-through estimator for the quantization operation. The final adversarial image is still saved in the original pixel space and is evaluated after the same input-side defense is applied.

The adaptive results show that the apparent protection from input transformations is brittle. Under adaptive optimization, ASR recovers to 82.5--93.0\% for JPEG compression, 97.0--98.0\% for Gaussian blur, and about 99.0\% for all bit-depth settings. This indicates that once the adversary optimizes using the same preprocessing operation as the defender, the adversarial image can be made robust to that transformation and still reach the malicious centroid after preprocessing.

Input transformations also introduce measurable degradation in benign utility. The utility drop ranges from 0.003 to 0.014 across the tested settings, with a stronger Gaussian blur causing the largest utility loss. Thus, simple input-side preprocessing also faces an unfavorable trade-off. It can suppress some non-adaptive attacks, but it degrades benign utility and can be largely bypassed by an adaptive adversary.
\section{Limitations and Future Directions}
\label{sec:limitations}

Our study has several limitations, and making them explicit improves both the technical clarity and the practical relevance of the paper.

\textbf{Defense practicality.}
There is a gap between strong theoretical defenses and realistic deployment constraints. Fully adapting the victim encoder or LLM may weaken the current attack, but in practice many users cannot afford such repairs due to training cost or operational constraints. This raises a broader research question about what defenses remain practical for resource-constrained PEFT ecosystems where only small components can be changed~\cite{hu2022lora}. We believe this question is more representative of real-world MLLM security than defenses that assume frequent full-model retraining.

\textbf{Attack objectives.}
Our current experiments focus on a fixed target phrase. This simplifies evaluation but does not exhaust the attack surface. Future work should study more realistic malicious objectives, including harmful instruction redirection, retrieval manipulation, multi-target backdoors, or conditional triggers tied to task context.

More broadly, future work should examine cross-modal latent space backdoors across stronger adaptive defenders and more realistic deployment settings. The present work demonstrates the existence of this threat, and the next step is to rigorously characterize the conditions under which it emerges and fails.
\section{Conclusion}
\label{sec:conclusion}

This work presents a cross-modal backdoor attack against connector-based MLLMs that exploits shared latent alignment to enable cross-modal backdoor activation. By poisoning only the lightweight connector, the adversary can establish a malicious latent region that remains activatable from other modalities through latent-space steering. Unlike conventional visible-trigger backdoors, our attack treats the shared latent representation space itself as the trigger carrier.

Extensive evaluations on representative connector-based MLLMs demonstrate that the attack achieves strong cross-modal activation while preserving benign utility, exhibiting low leakage on clean inputs and only minor deviation from the original connector parameters. These findings reveal that shared multimodal alignment creates a substantially broader attack surface than previously recognized.
This work suggests that risks introduced through one modality can propagate to others through aligned latent spaces. We hope these findings motivate stronger representation-level auditing and more comprehensive security evaluation for modular multimodal foundation models.

\section*{Ethical Considerations}
Our work evaluates the supply-chain risks of hijacking connector-based MLLMs, demonstrating how a poisoned lightweight module creates representation-level vulnerabilities that propagate across modalities. While exposing this emerging threat raises concerns about potential misuse, particularly for end users in high-stakes settings like content moderation or safety-critical perception, we argue that characterizing it in a controlled setting is crucial to help the research community and model providers develop stronger safeguards.

To minimize real-world risks, all experiments were conducted in isolated local environments using solely public models and datasets. We did not attack deployed services, utilize private user data, or upload poisoned artifacts to public model hubs. Furthermore, our evaluations exclusively employed harmless diagnostic sentences as target outputs. These choices ensure that we comprehensively study the security implications of connector poisoning without facilitating actual malicious behavior.

To balance reproducibility with security, we will responsibly release our experimental configurations, evaluation scripts, and sanitized examples, accompanied by strict usage restrictions. We will explicitly withhold ready-to-deploy poisoned checkpoints to avoid lowering the barrier to exploitation. This controlled release strategy ensures our artifacts exclusively empower independent validation and the development of defensive countermeasures.



%

\clearpage
\appendices

\twocolumn[{%
\begin{minipage}{\textwidth}
\centering
\captionof{table}{
Full results of the assessment of our attack.
Each cell reports Exact/Relaxed ASR (\%).
Each activation modality is evaluated in terms of ASR
under three different perturbation levels.
}
\label{tab:appendix-full-results}

\small
\setlength{\tabcolsep}{4.2pt}
\renewcommand{\arraystretch}{1.25}
\begin{tabular*}{\textwidth}{@{\extracolsep{\fill}}llccc ccc ccc@{}}
\toprule
\multirow{3}{*}{\textbf{Model}} 
& \multirow{3}{*}{\makecell{\textbf{Backdoor}\\\textbf{Modality}}}
& \multicolumn{9}{c}{\textbf{Activation Modality}} \\
\cmidrule(lr){3-11}
& 
& \multicolumn{3}{c}{\textbf{Image}} 
& \multicolumn{3}{c}{\textbf{Audio}} 
& \multicolumn{3}{c}{\textbf{Text}} \\
\cmidrule(lr){3-5} \cmidrule(lr){6-8} \cmidrule(l){9-11}
& 
& $8/255$ & $16/255$ & $32/255$ 
& $0.01$ & $0.05$ & $0.1$ 
& $0.01$ & $0.05$ & $0.1$ \\
\midrule

\multirow{3}{*}{PandaGPT} 
& Image 
& 92.5/92.5 & 96.5/96.5 & 99.5/99.5
& 0.0/0.0 & 98.0/98.5 & 99.2/99.5
& 21.0/21.0 & 98.5/98.5 & 99.4/99.5 \\

& Audio 
& 1.5/2.5 & 18.0/22.0 & 74.5/76.5
& 0.5/0.5 & 99.5/99.5 & 99.8/99.9
& 0.0/0.0 & 72.0/88.5 & 98.5/99.5 \\

& Text 
& 14.5/27.5 & 66.0/85.5 & 92.5/98.0
& 0.0/0.0 & 59.5/63.0 & 95.0/95.0
& 61.0/67.4 & 98.6/99.0 & 99.7/99.9 \\

\midrule

\multirow{3}{*}{NExT-GPT} 
& Image 
& 83.0/83.0 & 99.0/99.0 & 99.7/99.8
& 0.0/0.0 & 94.0/94.0 & 99.4/99.6
& 3.0/3.0 & 98.5/98.5 & 99.8/99.9 \\

& Audio 
& 10.0/12.0 & 59.0/64.0 & 96.0/96.0
& 7.0/8.0 & 98.8/99.0 & 99.6/99.9
& 0.0/6.7 & 99.0/99.0 & 99.9/99.9 \\

& Text 
& 33.0/36.0 & 87.0/90.0 & 99.0/99.0
& 0.0/0.0 & 97.0/98.0 & 99.8/99.8
& 82.0/82.0 & 97.0/97.5 & 99.9/99.9 \\

\midrule

\multirow{3}{*}{ChatBridge}
& Image
& 73.0/77.0 & 97.0/98.0 & 100.0/100.0
& 0.0/0.0 & 29.0/31.0 & 61.0/66.0
& N/A & N/A & N/A \\
& Audio
& 7.0/34.0 & 41.0/82.0 & 89.0/99.0
& 75.0/81.0 & 100.0/100.0 & 100.0/100.0
& N/A & N/A & N/A \\
& Text
& N/A & N/A & N/A
& N/A & N/A & N/A
& N/A & N/A & N/A \\
\bottomrule
\end{tabular*}

\end{minipage}
\par\vspace{\dbltextfloatsep}
}]

\section{Architecture of MLLMs}
\label{ap:preliminaries}

\textbf{Centralized LLM Dispatching.}
In the controller-style MLLM, the LLM does not directly consume non-text inputs. It operates over textual descriptions and orchestrates external expert modules for perception or generation. Systems such as HuggingGPT and Visual ChatGPT are representative examples of this design~\cite{shen2023hugginggpt,wu2023visualchatgpt}. However, this multi-step invocation of external tools introduces severe inference latency. Since multimodal inputs are processed through separate expert modules, the system may also lose fine-grained cross-modal alignment and exhibit weaker instruction-following consistency. Consequently, its mainstream adoption has significantly waned in favor of architectures that maintain continuous latent alignment. More importantly, These systems do not perform genuine latent-space multimodal fusion, placing them outside our target deployment setting.

\textbf{Native End-to-End Unification.}
In the native unified MLLM, diverse modalities are natively ingested and processed by a unified backbone from the ground up. Flagship commercial models, such as Google's Gemini~\cite{team2023gemini} and GPT-4o~\cite{openai2024gpt4o}, along with open-source efforts like Chameleon and Show-o~\cite{team2024chameleon,xie2024showo}, exemplify this end-to-end tokenization approach. While these native architectures eliminate the need for standalone multimodal bridges by building fusion directly into the backbone, their prohibitive from-scratch training costs render them largely inaccessible to standard developers, further driving the widespread adoption of the connector-based paradigms discussed above. These architectures likewise fall outside our threat model. Although they natively integrate multiple modalities, they lack an explicit, standalone component whose parameters can be directly manipulated to control modality interaction and redistributed independently.

Figure~\ref{fig:background_taxonomy} summarizes these three paradigms and highlights the one we target. Our attack focuses on the first family, which has become the predominant standard for open-source MLLMs. These models preserve fine-grained continuous signals, ensure high inference efficiency, and offer unparalleled plug-and-play convenience. However, this modularity introduces a critical vulnerability. The reason is inherently structural in that connector-based MLLMs expose a compact and trainable interface between encoders and the shared representation consumed by the LLM. This interface is explicit enough to be poisoned, yet influential enough to alter downstream behavior across tasks and modalities.

\section{Full Experimental Results}
\label{ap:full-results}

The complete evaluation results are reported in Table~\ref{tab:appendix-full-results}. For image inputs, we evaluate perturbation budgets in $\{8/255, 16/255, 32/255\}$. For audio and text inputs, we evaluate budgets in $\{0.01, 0.05, 0.1\}$. We use 500 optimization steps for PandaGPT and NExT-GPT, keeping activation lightweight while producing effective adversarial samples. For ChatBridge, we increase the number of steps to 12,000, as its query-based connector uses modality-specific learnable queries that can make cross-modal alignment more challenging than with MLP projectors. We evaluate 3,000 activation samples for each setting. 

The results show that at the intermediate perturbation budget of $0.05$, audio and text activation of an image-poisoned PandaGPT connector both achieve exact ASR of at least 98.0\%, while text activation of an audio-poisoned NExT-GPT connector reaches 99.0\%. These results already approach the success rates obtained at the largest evaluated budgets. High cross-modal effectiveness is therefore achievable with smaller perturbations in these settings, providing flexibility to balance perturbation magnitude and attack success when activating the same poisoned connector.

\end{document}